\documentclass[aps,prb,twocolumn,superscriptaddress]{revtex4-1}

\usepackage{graphicx}
\usepackage{amsmath,amsfonts,amssymb}
\usepackage[colorlinks=true,linkcolor=blue,citecolor=blue]{hyperref}

\usepackage{siunitx}
\usepackage{layouts}
\usepackage{booktabs}

\begin{document}

\title{Non-equilibrium properties of graphene probed by superconducting tunnel spectroscopy}


\author{Simon Zihlmann}
\email{simon.zihlmann@unibas.ch}
\affiliation{Department of Physics, University of Basel, Klingelbergstrasse 82, CH-4056 Basel, Switzerland}

\author{P\'eter Makk}
\email{peter.makk@mail.bme.hu}
\affiliation{Department of Physics, University of Basel, Klingelbergstrasse 82, CH-4056 Basel, Switzerland}
\affiliation{Department of Physics, Budapest University of Technology and Economics and Nanoelectronics 'Momentum' Research Group of the Hungarian Academy of Sciences, Budafoki ut 8, 1111 Budapest, Hungary}

\author{Sebasti\'an Castilla}
\affiliation{Department of Physics, University of Basel, Klingelbergstrasse 82, CH-4056 Basel, Switzerland}

\author{J\"{o}rg Gramich}
\affiliation{Department of Physics, University of Basel, Klingelbergstrasse 82, CH-4056 Basel, Switzerland}

\author{Kishan Thodkar}
\affiliation{Department of Physics, University of Basel, Klingelbergstrasse 82, CH-4056 Basel, Switzerland}

\author{Sabina Caneva}
\affiliation{Department of Engineering, University of Cambridge, 9 JJ Thomson Avenue, Cambridge CB3 0FA, United Kingdom}

\author{Ruizhi Wang}
\affiliation{Department of Engineering, University of Cambridge, 9 JJ Thomson Avenue, Cambridge CB3 0FA, United Kingdom}

\author{Stephan Hofmann}
\affiliation{Department of Engineering, University of Cambridge, 9 JJ Thomson Avenue, Cambridge CB3 0FA, United Kingdom}

\author{Christian Sch\"onenberger}
\affiliation{Department of Physics, University of Basel, Klingelbergstrasse 82, CH-4056 Basel, Switzerland}

\date{\today}
\begin{abstract}
We report on non-equilibrium properties of graphene probed by superconducting tunnel spectroscopy. A hexagonal boron nitride (hBN) tunnel barrier in combination with a superconducting Pb contact is used to extract the local energy distribution function of the quasiparticles in graphene samples in different transport regimes. In the cases where the energy distribution function resembles a Fermi-Dirac distribution, the local electron temperature can directly be accessed. This allows us to study the cooling mechanisms of hot electrons in graphene. In the case of long samples (device length $L$ much larger than the electron-phonon scattering length $l_{e-ph}$), cooling through acoustic phonons is dominant. We find a cross-over from the dirty limit with a power law $~T^3$ at low temperature to the clean limit at higher temperatures with a power law $~T^4$ and a deformation potential of \SI{13.3}{\eV}. For shorter samples, where $L$ is smaller than $l_{e-ph}$ but larger than the electron-electron scattering length $l_{e-e}$, the well-known cooling through electron out-diffusion is found. Interestingly, we find strong indications of an enhanced Lorenz number in graphene. We also find evidence of a non-Fermi-Dirac distribution function, which is a result of non-interacting quasiparticles in very short samples.
\end{abstract}

\maketitle

\section{Motivation/Introduction}
\label{sec:Introduction}
Graphene is particularly interesting for non-equilibrium transport studies, since its 2D nature results in deviations from the heavily studied 3D bulk case. Since optical phonon energies are quite large in graphene \cite{2011_DasSarma}, they can be neglected at low temperatures, and acoustic phonon cooling can lead to different regimes depending on phonon wavelength and electron mean free path \cite{2010_Viljas, 2012_Chen}. Moreover, due to the low density of states reduced screening can alter the strength of electron cooling \cite{2008_Mueller}. Finally special non-equilibrium regimes can appear like the supercollision regime \cite{2012_Song, 2012_Betz, 2012_Graham, 2014_Laitinen} or the Dirac fluid regime \cite{2016_Crossno}.

\begin{figure}[htbp]
	\includegraphics[scale=1]{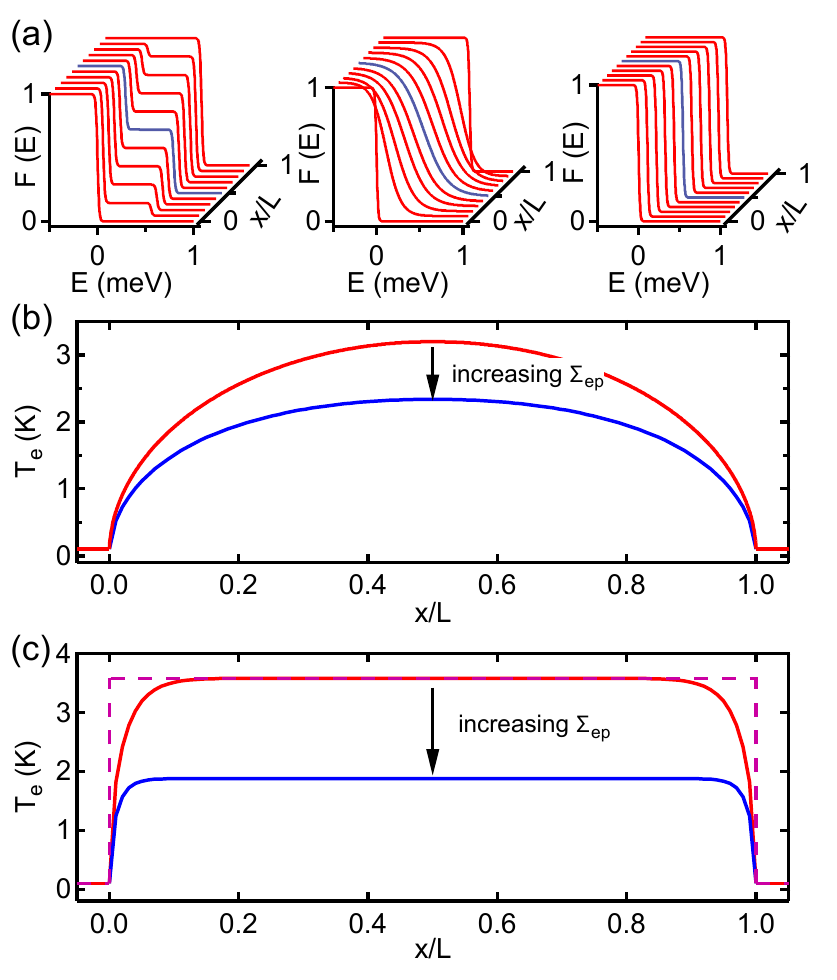}
	\caption{\label{fig:supra_T_profile}\textbf{(a) Non-equilibrium distribution functions:} \textbf{(Left)} The distribution function in the absence of inelastic scattering, where energy relaxation only occurs in the reservoirs. \textbf{(Middle)} The distribution function in the case of strong quasiparticle scattering where local thermal equilibrium is achieved, but phonon scattering is negligible. \textbf{(Right)} In the case of very strong phonon scattering, the quasiparticles thermalize with the phonons and a constant temperature along the channel is found. \textbf{(b)} \textbf{Temperature profile in the hot electron regime:} $T_e(x)$ is obtained by numerically solving the heat transfer equation~\ref{eq:theory_heat_equation_num}. The dimensions of device A are used and the effect of increasing electron-phonon coupling is shown. \textbf{(c) Temperature profile in the phonon cooled regime:} The dimensions of device C are used for two different electron phonon coupling strength. The effect of the hot electron out-diffusion is only seen close to the contacts where the red solid line deviates from the dashed purple line that neglects cooling through electron out-diffusion. The reservoir temperature is assumed to be \SI{100}{\milli\kelvin} for both cases.} 
\end{figure}

In general, different temperature profiles and distribution functions arise depending on the sample size and the scattering mechanisms involved \cite{2013_Heikkilae}. Usually the sample is connected to two normal metal contacts (N1 and N2) that can be biased to different electrochemical potentials. Such a bias $U$ will lead to Joule heating, that heats the electron system.  This can lead to non-uniform temperature distributions and in some cases the electronic distribution function even deviates from a Fermi-Dirac distribution.
If only elastic scattering among charge carries happens (device length $L$ and device width $W$ are much shorter than the electron-electron scattering length $l_{e-e}$ and electron-phonon scattering length $l_{e-ph}$) the distribution function will take the form of a double-step function as shown in Fig.~\ref{fig:supra_T_profile}~(a, left). Here we assume, that the electrodes are ideal reservoirs that absorb all incoming quasiparticles and emit quasiparticles with an energy distribution given by their own Fermi-Dirac distribution\cite{1957_Landauer}. If the device is larger, such that $l_{e-e}<L,W < l_{e-ph}$, electron-electron scattering leads to Joule heating dissipated only into the electron system. This regime is also called hot electron regime and the corresponding energy distribution function inside the graphene is shown in Fig.~\ref{fig:supra_T_profile}~(a, middle). The energy distribution function is well described by a Fermi-Dirac distribution with an effective electron temperature that depends on the position in x-direction (local thermal equilibrium), with a maximum in temperature in the middle of the sample (see Fig.~\ref{fig:supra_T_profile}(b)). In very long graphene channels, where $L>l_{e-ph}$, most of the Joule power is dissipated to the lattice through phonon emission. This leads to a constant electron temperature along the channel with a Fermi-Dirac distribution function of the quasiparticles, see Figs.~\ref{fig:supra_T_profile}(a, right) and (c). Here, the electron-phonon coupling in graphene is the bottleneck in cooling to the substrate since the acoustic phonons in graphene are very well coupled to the SiO$_2$ substrate \cite{2009_Chen_a}.

Noise measurements have proven to be quite powerful to access the electronic temperature of nanostructures \cite{1996_Steinbach, 1997_Henny, 1999_Henny} and in particular of graphene \cite{2012_Betz, 2013_Fong, 2014_Laitinen, 2015_Brunel, 2016_McKitterick}. In certain regimes the electronic temperature can also be obtained from supercurrent \cite{2013_Borzenets} and from quantum Hall measurements \cite{2012_Baker}. A different and more direct approach is to study the non-equilibrium energy distribution function itself, which can be done using superconducting tunnel spectroscopy \cite{1996_Pothier}. This method was introduced by Pothier et al. on metallic nanowires \cite{1997_Pothier} and was also used to study carbon nanotubes \cite{2009_Chen}.

In this paper we use this latter method to study the non-equilibrium distribution function in graphene under different biases and for different sample geometries. We use hexagonal boron nitride as a tunnel barrier and Pb as a large gap superconductor to increase our spectroscopic range. Our results are compared with simple heat equation models containing both electron and phonon cooling. For short samples we find that the distribution functions are well described by a Fermi-Dirac distribution with effective temperatures that are in agreement with the expectations for the hot electron regime, however, with an increased effective Lorenz number. For larger samples phonon cooling dominates and a crossover from the clean to the dirty limit as a function of heating power is observed. We extract the value of the electron-phonon coupling for both regimes.

The paper is organized as follows. We present our thermal model in Sec.~\ref{sec:Thermal} that can describe all experimental cases. Our measurements and extraction methods are shown in Sec.~\ref{sec:Methods}. Sec.~\ref{subsec:hot_electron} describes our results in the hot electron regime, whereas Sec.~\ref{subsec:phonon_cooling} discusses the results in the phonon cooled regime. Sec.~\ref{subsec:double_step} shows evidence of a double-step distribution function, which is followed by a general conclusion in Sec.~\ref{sec:Conclusion}.

\section{Thermal model}
\label{sec:Thermal}
If a local temperature can be defined (hot electron and phonon cooled regime), thermal transport can be described by the continuity equation for heat, which relates the difference of the change of energy density over time and the gradient of the heat current to the local sources and sinks:
\begin{equation}
	\label{eq:theory_heat_continuits}
	\rho_M c_p\frac{\partial T_e(x,t)}{\partial t} - \frac{\partial}{\partial x}\left(\kappa \frac{\partial T_e(x, t)}{\partial x}\right) = P -P_{ph},
\end{equation}
where $\rho_M=$~\SI{7.6e-7}{\kilo\gram\per\square\metre} is the mass density of graphene, $c_p$ is the specific heat capacity, $T_e(x,t)$ is the local electron temperature, $\kappa$ is the heat conductivity, $P$ is the Joule heating power per unit area and $P_{ph}$ is the phonon cooling power per unit area. In steady state, $\frac{\partial T(x,t)}{\partial t} = 0$, and using the Joule heating $P = U^2/R$ and the device dimensions $W,L$ as defined in Fig.~\ref{fig:supra_device}~(a) one arrives at
\begin{equation}
	\label{eq:theory_heat_equation}
\begin{aligned}
	\frac{U^2}{R} ={} & -LW\frac{\partial}{\partial x}\left(\kappa_{WF}(x)\frac{\partial T_e(x)}{\partial x}\right) +\\
	& LW\cdot\Sigma_{ep}\left(T_e(x)^\delta - T_0^\delta\right),
\end{aligned}
\end{equation}
where the Joule heating on the left side is balanced by cooling through electron diffusion (first term on the right) and cooling through electron-phonon coupling (second term on the right). Here, $U$ is the bias applied across the device resistance $R$ and $T_0$ is the phonon temperature. Furthermore, the electron cooling is connected to the electrical conductivity through the Wiedemann-Franz law \cite{1853_Wiedemann, 2013_Heikkilae} $\kappa_{WF}(x) = \mathcal{L}_0 T_e(x)L/(WR)$, where $\mathcal{L}_0 = \frac{\pi^2k_B^2}{3e^2}$ is the Lorenz number. The electron-phonon cooling can be parametrized through the coupling constant $\Sigma_{ep}$ and the exponent $\delta$, which can depend on temperature and device properties \cite{2012_Chen}. If the explicit form of the Wiedemann-Franz law is plugged into Eq.~\ref{eq:theory_heat_equation}, one arrives at the following relation
\begin{equation}
	\label{eq:theory_heat_equation_num}
	 \begin{aligned}
	\frac{\partial^2T_e(x)}{\partial x^2} ={}& \frac{1}{T_e(x)} \left[-\frac{U^2}{L^2\mathcal{L}_0} + \right. \\
	& \left. \frac{RW}{L\mathcal{L}_0}\Sigma_{ep}\left(T_e(x)^\delta - T_0^\delta\right) - \left(\frac{\partial T_e(x)}{\partial x}\right)^2 \right],
\end{aligned}
\end{equation}
which can be used to numerically solve for $T_e(x)$.

In the absence of phonon cooling ($\Sigma_{ep}\rightarrow 0$), the above equation has the simple solution
\begin{equation}
	\label{eq:theory_hot_electron_T}
	T_{e}(x) = \sqrt{T_0^2 + \frac{x}{L}\left(1-\frac{x}{L}\right)\frac{U^2}{\mathcal{L}_0}},
\end{equation}
where the temperature at the electrodes $T_e(x=0) = T_e(x=L) = T_0$ is used as a boundary condition. The local electron temperature is shown in Fig.~\ref{fig:supra_T_profile}~(b).

On the other hand if electron cooling is negligible (e.g. very large thermal contact resistance or superconducting contact materials that suppress cooling through electron diffusion), the electron temperature is position independent and given by:
\begin{equation}
	\label{eq:theory_phonon_T}
	T_e(x) = \sqrt[\delta]{\frac{U^2}{RLW\Sigma_{ep}} - T_0^\delta}.
\end{equation}
In this case, there will be a discontinuity of the temperature across the contact to the graphene, see Fig.~\ref{fig:supra_T_profile} (c). In a simple case where $T_0\rightarrow$~\SI{0}{\kelvin} and assuming $\delta = 4$ (clean limit, see below), the transition between electron cooling and phonon cooling happens at a bias voltage $U_{\Sigma} = \mathcal{L}_0 / \sqrt{4RLW\Sigma_{ep}}$ \cite{2012_Betz_a}. For typical devices with dimensions on the order of \si{\micro\metre}, device resistance of \si{\kilo\ohm} and an electron-phonon coupling of around \SI{30}{\milli\watt\kelvin^{-4}\metre^{-2}} \cite{2016_McKitterick} the crossover voltage is on the order of \si{\milli\volt} .

Phonon cooling at low temperature is dominated by acoustic phonons since optical phonon energies are on the order of \SI{100}{\milli\eV} \cite{2004_Piscanec}. In the limit of $T_e\gg T_0$, the cooling power takes the approximate form of $P\approx\Sigma T_e^\delta$, that allows to extract the power $\delta$ and electron-phonon coupling $\Sigma$ easily. Here we would also like to note that we work well below the Bloch-Gr\"uneisen temperature $T_{BG} = 2s\hbar\sqrt{\pi n}/k_B$, where $s=$~\SI{2e4}{\metre\per\second} is the speed of sound in graphene, $n$ is the carrier density and $k_B$ is Boltzmann's constant. For a reasonable doping of  $n\sim$~\SI{1e12}{\per\square\centi\metre}, $T_{BG}$ is estimated to be around \SI{50}{\kelvin}. At temperatures below $T_{BG}$ the phase-space of electron-phonon scattering is greatly reduced and only small angle scattering is possible \cite{2010_Efetov}.

The electron-phonon coupling can strongly be modified by electronic disorder if the wavelength of the thermal phonons becomes comparable to (or longer than) the electronic mean free path \cite{2012_Chen}. This condition results in two regimes (even below $T_{BG}$); the clean limit where the mean free path is much longer than the phonon wavelength and the dirty limit where the mean free path is much shorter than the phonon wavelength.\\
In the clean limit and assuming a weak screening, the total cooling power due to electron-phonon interaction is given as \cite{2012_Chen}
\begin{equation}
	\label{eq:theory_heat_phonon_T4}
	P(T_e, T_0) = A\Sigma_1(T^4_e - T^4_0),\quad \Sigma_1 = \frac{\pi^2D^2\mid E_F\mid k_B^4}{15\rho_M\hbar^5v_F^3s^3}.
\end{equation}
Here, $D$ is the deformation potential that characterizes the strength of the electron-phonon coupling, and the other symbols are defined above. In contrast to that, the cooling power due to the electron-phonon interaction in the dirty limit is \cite{2012_Chen}
\begin{equation}
	\label{eq:theory_heat_phonon_T3}
	P(T_e, T_0) = A\Sigma_2(T^3_e - T^3_0),\quad \Sigma_2 = \frac{2\zeta(3)D^2\mid E_F\mid k_B^3}{\pi^2\rho_M\hbar^4v_F^3s^2l_{mfp}}.
\end{equation}
Here, $\zeta(n)$ is the Riemann zeta function with $\zeta(3)\approx1.2$.  The crossover between these two regimes is characterized by $T_x$, at which the cooling power of the clean and dirty limit is equal. This temperature is given by
\begin{equation}
	\label{eq:theory_heat_phonon_T_dis}
	T_x = \frac{30\hbar s\zeta(3)}{\pi^4 k_B l_{mfp}}.
\end{equation}
Graphene samples on SiO$_2$ substrates generally show a mean free path of the order of \SI{30}{\nano\metre}, which leads to a crossover temperature $T_x\sim$~\SI{1}{\kelvin}. Experimentally, the different cooling regimes can be accessed by varying the electron temperature and the heating power (e.g. Joule heating) applied to the electronic system. 

Having introduced a thermal model describing a graphene channel driven out-of-equilibrium, we now proceed to the methods and our results that show clear evidence for all these regimes introduced above.


\section{Methods}
\label{sec:Methods}

\subsection{Sample fabrication}
\begin{figure}[htbp]
	\includegraphics[scale=1]{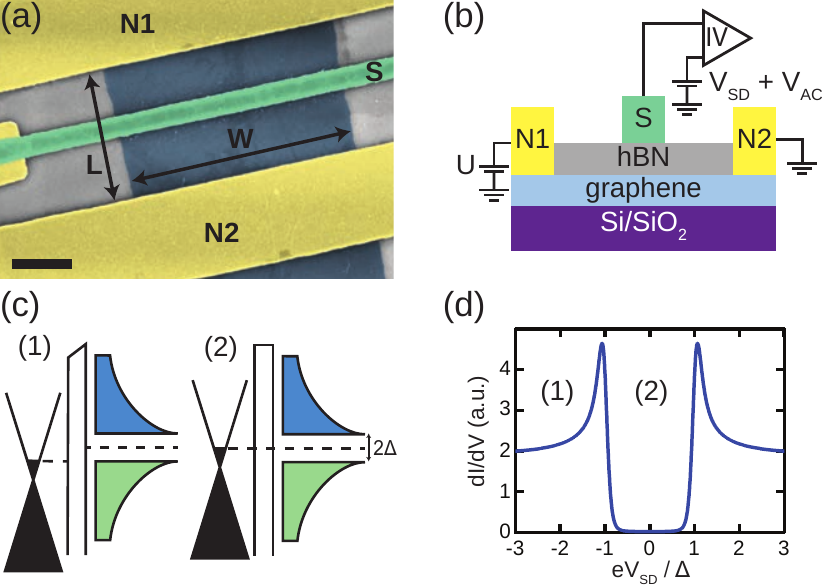}
	\caption{\label{fig:supra_device}\textbf{Device and working principle of superconducting tunnel spectroscopy:} \textbf{(a)} shows a false colour scanning electron micrograph of a typical sample. The superconducting Pd/Pb/In electrode is labelled by S and the normal Cr/Au contacts are labelled with N1 and N2, respectively. The scale bar is \SI{1}{\micro\metre}. \textbf{(b)} shows a cross section of a typical device with the measurement setup indicated. \textbf{(c)} and \textbf{(d)} show the working principle of superconducting tunnel spectroscopy with the energy diagrams in \textbf{(c)} and the resulting  differential conductance in \textbf{(d)}. Current can only flow if the bias $V_{SD}$ across the tunnel junction is larger than $\Delta/e$ case (1), otherwise it is suppressed due to the gapped DoS, case (2).}
\end{figure}

A false colour micrograph of a typical device is shown in Fig.~\ref{fig:supra_device}~(a) with a cross section in (b). It consists of a graphene channel of length $L$ and width $W$, which is connected to two normal contacts N1 and N2 that act as ideal reservoirs: all incoming quasiparticles are absorbed and the emitted quasiparticles obey a Fermi-Dirac distribution of the respective reservoir. In the middle of the graphene channel, a superconducting electrode S is tunnel coupled to the graphene. We employ chemical vapour deposited (CVD) single or two layer hBN films as tunnel barriers, covering the full sample.

Here, we used CVD graphene grown in-house following the recipe described in Ref.~\cite{2017_Thodkar}. After the transfer from the growth substrate to a Si/SiO$_2$ wafer, the graphene was structured by e-beam lithography and reactive ion etching into the desired shape. The CVD hBN layer was transferred after a thermal annealing in forming gas. Commercial hBN obtained from Graphene Supermarket \cite{graphene_supermarket} was used for devices A and D, whereas hBN from the Hofmann group \cite{2016_Caneva} was used for device B and C. In the case of devices A and D a two layer CVD hBN was used (sequential transfer of two single layer hBN). Single layer hBN was used for devices B and C. An overview of all devices can be found in Tab.~\ref{tab:supra_devices}. In a next step, the normal contacts were fabricated. Quasi 1-dimensional Cr/Au (\SI{10}{nm}/\SI{50}{nm}) side contacts were achieved by a short plasma etching before the metal deposition. The CHF$_3$ based plasma removed the hBN, as well as partially the graphene. It turned out that these quasi 1-dimensional side contacts are less reproducible than the 1-dimensional contacts to "bulk" hBN/Gr/hBN vdW-heterostructures. A significant increase in the number of working contacts was achieved by redeveloping the PMMA mask after the CHF$_3$ plasma. We attribute this to the fact that the overlap of the metal with the graphene channel is increased in this case. In a last step, the superconducting electrode was deposited. Here, we used an optimized three layer structure consisting of \SI{5}{\nano\metre} Pd as wetting layer, \SI{110}{\nano\metre} of Pb and \SI{20}{\nano\metre} of In as a capping layer \cite{2016_Gramich}. The Pd sticking layer was used to avoid oxidation at the hBN/Pb interface\cite{2013_Li} and the In capping layer was used to prevent oxidation from the top. Pb was chosen since its large superconducting gap allows thermometry up to several kelvins.

\begin{table}[t]
	\centering
	\renewcommand{\arraystretch}{1.2}
	\caption{\label{tab:supra_devices}\textbf{Overview of all devices:} $L$ and $W$ are specified in Fig.~\ref{fig:supra_device}, $A_T$ denotes the area of the tunnel contact and $R_TA_T$ specifies the tunnel resistance area product of the tunnel contact. Here we differentiate between commercially available CVD hBN (comm.) obtained from Graphene Supermarket\cite{graphene_supermarket} and self-grown CVD hBN (collab.) \cite{2016_Caneva}.}
	\begin{tabular}{lllll>{\centering\arraybackslash}m{2cm}}
    \toprule[1.5pt]
		~	& $L$ (\si{\micro\metre})	& $W$ (\si{\micro\metre}) & $R_TA_T$ (\si{\ohm\square\micro\metre}) & $A_T$ (\si{\square\micro\metre}) & hBN source\\ \hline
		A	&3.4	&2	&\num{90e3}				& 0.7		& 2-layer, comm.\\
		B 	&1.3	&4  &\num{470e3}			& 1.4		& 1-layer, collab.\\
		C 	&100	&6	&$\leq$\num{40e3}	&	2.1		& 1-layer, collab.\\
		D	&2.5	&1	&\num{1.8e6}			&	0.35	& 2-layer, comm.\\
    \bottomrule[1.5pt]
	\end{tabular}
\end{table}

All measurements shown in here were carried out in a dilution fridge at low temperatures of \SI{\sim 50}{\milli\kelvin}. The electrical measurement scheme is shown in Fig.~\ref{fig:supra_device}(b). For the measurement of the differential conductance a standard low-frequency lock-in technique was used. The source-drain voltage consists of a DC and an AC part, which are $V_{SD}$ and $V_{AC}$ respectively. Both voltages were applied by shifting the ground potential of the I-V converter hooked up to the S contact, whereas the heating voltage was directly applied on contact N1, while N2 was grounded.  Very low AC excitation voltages on the order of $k_BT/e$ have been used in order to maximize the energy resolution. The carrier density was tuned by applying $V_{BG}$ on the doped Si-substrate.

\subsection{Working principle of superconducting tunnel spectroscopy}
To obtain the distribution function and the effective temperature during non-equilibrium conditions we have measured the tunnelling conductance from the superconducting electrode, while the channel was driven out-of equilibrium via a large bias. Here we summarize shortly how the distribution function and the effective temperature was obtained. Further details can be found in the Appendix~\ref{sec:deconvolution}.

The differential conductance of a superconductor (S)/insulator (I)/graphene (gr) junction is given by\cite{2013_Heikkilae, 1996_Pothier}:
\begin{equation}
	\label{eq:supra_SIN_dIdV}
\begin{aligned}
	\frac{\mathrm{d}I(V)}{\mathrm{d}V} = {} & \frac{1}{R_T}\int_{-\infty}^{+\infty} \mathrm{d}E \cdot \mathrm{n_{gr}}(E)\\
	& \cdot\frac{\mathrm{d}n_s(E-eV)}{\mathrm{d}E}\cdot\left(f_S(E-eV) - f_{gr}(E)\right),
\end{aligned}
\end{equation}
where $R_T$ is the tunnel junction resistance, $n_s(E)$ is the superconducting density of states (DoS) with an energy gap of $\Delta$, $n_{gr}(E)$ is the  DoS of graphene and $f_s(E)$, $f_{gr}(E)$ are the energy distribution functions in the superconductor and in the graphene, respectively \cite{2013_Heikkilae}. An energy diagram of the tunnelling process is shown in Fig.~\ref{fig:supra_device}~(c) next to the differential conductance, see Fig.~\ref{fig:supra_device}~(d).

The energy distribution function in graphene $f_{gr}(E)$ can be obtained by a deconvolution of the measured $dI/dV$ using Eq.~\ref{eq:supra_SIN_dIdV}. In order to do so, $n_{gr}$, $n_s$ and $f_s$ need to be known. The density of states of the graphene $n_{gr}(E)$ can be assumed constant for small biases on the \si{\milli\eV}--scale for the large dopings we will use. If $k_BT\ll\Delta$, then the energy distribution function in the superconductor is well described by the Heavyside function $\Theta(E-eV)$ instead of a Fermi-Dirac distribution $F(E-eV)$. In addition, if the energy distribution function in graphene is a Heavyside function (i.e. a very cold Fermi gas), then the $dI/dV$ is directly proportional to the DoS ($n_s$) of the superconductor. Therefore, the $dI/dV$ measured at the lowest temperature with zero heating bias $U$ directly resembles $n_s$ that can be used for the numerical deconvolution.

\begin{figure*}[tb]
	\includegraphics[scale=1]{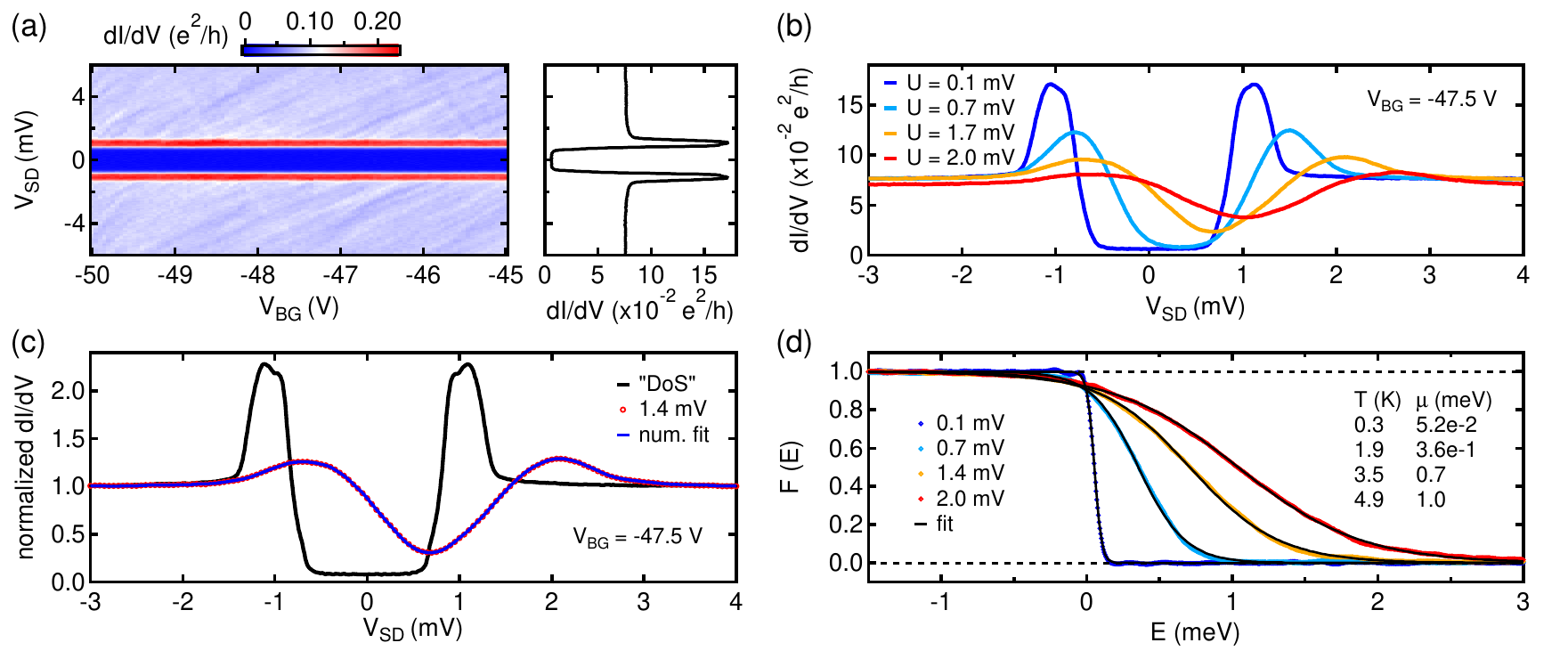}
	\caption{\label{fig:supra_measurement_principle}\textbf{Extraction of the distribution function:} \textbf{(a)} shows the differential conductance measured through the superconducting electrode to the graphene as a function of gate voltage $V_{BG}$ and bias across the hBN tunnel barrier $V_{SD}$. A pronounced superconducting gap of the Pd/Pb/In electrode is clearly observed. In addition some resonances tuned by $V_{BG}$ and $V_{SD}$ are visible outside of the gap, which are attributed to UCFs. In order to remove those resonances, an averaging over \SI{5}{\volt} in $V_{BG}$ is performed and the average is shown on the right. All subsequent differential conductance traces are averaged over \SI{5}{\volt} around the indicated gate voltage. \textbf{(b)} shows the differential conductance for different values of $U$ applied across the graphene flake to drive it out-of-equilibrium. A clear broadening of the gap due to heating is observed, while the position of the gap shifts by roughly $U/2$. \textbf{(c)} By using the lowest T and $U=$~\SI{0}{\volt} measurement as the density of states (DoS) of the superconductor, the differential conductance at $U\neq$~\SI{0}{\volt} can be used to numerically deconvolve the energy distribution function. \textbf{(d)} shows the numerically extracted energy distribution functions from the three traces shown in \textbf{(b)}. They all resemble a Fermi-Dirac distribution and therefore the electron temperature and the electrochemical potential can be extracted.}
\end{figure*}

Fig.~\ref{fig:supra_measurement_principle}~(a) shows the differential conductance measured from the superconductor to the graphene with N1 and N2 grounded (\SI{0}{V}) as a function of the spectroscopy bias $V_{SD}$ and back gate voltage $V_{BG}$. A clear superconducting gap is observed. Since there are some resonances tuned by both $V_{BG}$ and $V_{SD}$, an averaging over \SI{5}{\volt} in $V_{BG}$ is performed that is shown on the right. These resonances most probably originate from universal conductance fluctuations (UCFs). As stated above, this measurement resembles the DoS of the superconductor. A zoom in of the same measurement is shown in Fig.~\ref{fig:supra_measurement_principle}~(c). It is obvious that this DoS cannot be described by a standard BCS or a Dynes DoS \cite{2004_Tinkham, 1978_Dynes} as expected for a superconductor. The deviation might be due to the averaging that is needed to get rid of the fluctuations present in (a). This averaging then leads to a much broader peak at the gap edge than predicted by a BCS or a Dynes density of states \cite{2004_Tinkham, 1978_Dynes}. We, therefore, use the lowest temperature and zero heating bias $U$ measurement as the DoS of the superconductor.

The presence of a finite heating bias $U$ across the graphene channel (applied between N1 and N2, see Fig.~\ref{fig:supra_device}~(b)) drives the electronic system out-of-equilibrium. Fig.~\ref{fig:supra_measurement_principle}~(b) shows the tunneling differential conductance at several values of heating bias $U$. Two main changes can be observed: First, the superconducting gap smears out and second, the position of the superconducting gap shifts in $V_{SD}$. The smearing can be explained by a higher electron temperature and the shift in bias is just due to the linear voltage drop of $U$ along the graphene channel that shifts the electrochemical potential below the superconductor by $eU/2$.

Fig.~\ref{fig:supra_measurement_principle}~(c) shows the DoS of the superconductor and the tunneling differential conductance at $U=$~\SI{1.4}{\milli\volt}. In a numerical deconvolution, the energy distribution function of the graphene at finite heating bias $U$ can be extracted. To do so, a reasonable guess of the energy distribution function is assumed and according to Eq.~\ref{eq:supra_SIN_dIdV} the resulting tunneling differential conductance is calculated. The calculated differential conductance is then compared to the measurement and based on the differences, the guess of the energy distribution function is adjusted. This procedure is repeated until it matches the measured $dI/dV$, see Fig.~\ref{fig:supra_measurement_principle}~(c). Details about this numerical deconvolution can be found in appendix~\ref{sec:deconvolution}.

The corresponding energy distribution functions to the differential conductance measurements in Fig.~\ref{fig:supra_measurement_principle}~(b) are shown in Fig.~\ref{fig:supra_measurement_principle}~(d). In this case (device B) the energy distribution functions resemble a Fermi-Dirac distribution, which is parametrized by the electron temperature $T_e$ and the electrochemical potential $\mu$. These two parameters were extracted by fitting a Fermi-Dirac distribution function to the numerically extracted energy distribution functions.

We have verified the method by measuring the tunneling differential conductance at different bath temperatures. The extracted distribution functions were fitted with a Fermi-Dirac function and the extracted temperatures were compared with the bath temperature. We obtained reasonable agreement at higher temperatures, whereas below 100$\,$mK the extraction was limited by the fact that the used DoS was measured at a similar temperature. Therefore, the DoS already contains a small temperature broadening. This broadening becomes relevant at such low temperatures but can be fully neglected a higher temperatures. Since the electronic temperatures investigated will be larger than 100$\,$mK this will not limit our resolution. These measurements and a detailed discussion are given in Appendix~\ref{sec:Temp_dependence}.

\section{Results}
\label{sec:Results}

\subsection{Hot electron regime}
\label{subsec:hot_electron}

\begin{figure}[htbp]
	\includegraphics[scale=1]{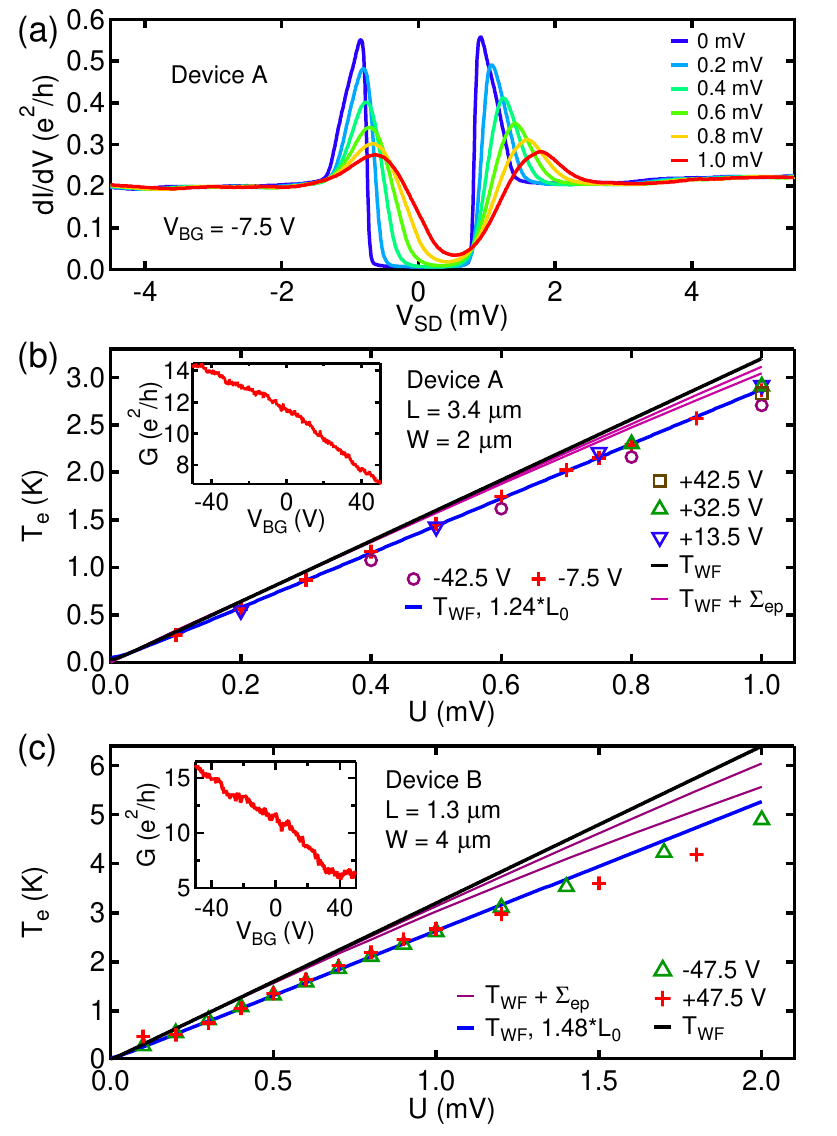}
	\caption{\label{fig:supra_hot_electron1}\textbf{(a)-(b) Device A in the hot electron regime:} \textbf{(a)} shows the differential conductance measured through the superconducting electrode to the graphene for different values of heating bias $U$ at a gate voltage of \SI{-7.5}{\volt}. \textbf{(b)} shows the extracted electron temperature from fitting a Fermi-Dirac distribution to the numerically extracted distribution function. The electron temperature increases nearly linear with applied bias as expected for a dominating cooling mechanism due to electron out-diffusion. The cooling mechanism is independent of gate voltage. The inset shows the two-terminal conductance through the graphene from N1 to N2 as a function of gate voltage $V_{BG}$. \textbf{(c) Device B in the hot electron regime:} The extracted electron temperature for two different gate voltages is shown. The extracted temperature deviates significantly from a linear dependence at higher bias, which is attributed to an additional cooling by phonons on top of the increased Lorenz number. The inset shows the two-terminal conductance through the graphene from N1 to N2 as a function of gate voltage $V_{BG}$ }
\end{figure}

Fig.~\ref{fig:supra_hot_electron1}~(a) shows the tunnelling differential conductance for several values of heating bias $U$ for device A. An increased $U$ leads to a smearing of the sharp superconducting gap and the middle of the gap is shifted by $U/2$ since the tunnel probe is located in the middle of the sample. The extracted electron temperature is shown in Fig.~\ref{fig:supra_hot_electron1}~(b) as function of heating bias $U$ for several values of back gate voltage $V_{BG}$. It can be seen that $T_e$ depends linearly on $U$, as expected for the hot-electron regime (for $T \gg T_0$). The inset in (b) shows the gate dependence of the graphene conductance measured from N1 to N2. While the graphene resistance is tuned by roughly a factor of two (by changing the charge carrier density by \SI{\sim 7e12}{\per\square\centi\metre}), the resulting electron temperature is independent of the graphene resistance and charge carrier concentration.

Using Eq.~\ref{eq:theory_hot_electron_T} the electron temperature profile can be calculated analytically and the expected $T_e$ at the location of the superconducting probe electrode is shown as solid black line in Fig.~\ref{fig:supra_hot_electron1}~(b). We note here, that this line is not a fit, and does not contain any free parameters. However all extracted values for $T_e$ fall below the expected value. We will come back to the discussion of this deviation below.

Similar results have been obtained for device B, which are shown in Fig.~\ref{fig:supra_hot_electron1} (c). The extracted electron temperatures of device B are also independent of the gate voltage. Here, the graphene resistance changes by a factor of three while changing the charge carrier density by \SI{\sim 7e12}{\per\square\centi\metre}. This again confirms the negligible role of electrical contact resistance. Here, the dependence of the temperature on heating bias $U$ can be divided into two qualitatively different regimes. For $U\leq$~\SI{1}{\milli\volt}, a linear dependence similar to device A is observed. Again, the extracted values for $T_e$ are smaller than calculated by Eq.~\ref{eq:theory_hot_electron_T} (solid black line).  For $U>$~\SI{1}{\milli\volt}, the extracted electron temperature is much lower than expected and becomes sublinear. This change in dependence could be explained by the onset of electron-phonon cooling which reduces the electron temperature below the expected value. 

Now we discuss the possible reasons for the reduced temperature compared to the expectations based on the hot electron regime. 

If a large contact resistance would be present, a substantial part of the heating bias would drop on that and a bias smaller than U would drive the graphene out-of-equilibrium. The ratio of the voltage dropping on the graphene and on the contact resistance would depend on the gate voltage as the graphene resistance is gate-tunable. Therefore, different temperatures for the same $U$ would be expected for different gate voltages. However, the measurements show that neither the electrical contact nor the charge carrier density plays a significant role and we therefore rule out a significant contact resistance.

Second, a finite contact resistance between the graphene and the normal metal reservoirs could lead to a thermal contact resistance as well. The presence of a thermal contact resistance would lead to a larger electron temperatures in the graphene as the cooling would be less efficient. Similar arguments hold for reservoirs that are at an elevated temperature. Both effects would lead to higher electron temperatures and are therefore ruled out.

In principle cooling through the superconducting electrode could also occur. However, first, the contact resistance is on the order of \SI{100}{\kilo\ohm}, which is roughly 100 times larger than the total device resistance. Therefore, only a correction on the order of \SI{1}{\percent} can be expected. Second, the reduced density of states in the superconductor at the Fermi energy efficiently suppresses cooling through electron out diffusion \cite{2013_Borzenets, 2013_Fong}.

In the hot electron regime, $T_e(x)$ is described by a pseudoparabolic profile. Obviously a superconducting electrode with finite width will not only probe the highest temperature in the middle, but will also probe lower temperatures off-centre. In order to estimate this, the width of the superconducting electrode (\SI{\leq 400}{\nano\metre}) has to be compared to the device length (\SI{3.4}{\micro\metre} for device A and \SI{1.3}{\micro\metre} for device B). Even though the width of the superconducting electrode is a considerable fraction of the device length for device B, its influence is estimated to be smaller than \SI{1.6}{\percent}, therefore, this effect is too small to explain the deviation from the expected electron temperature.

Obviously, cooling through phonons lowers the electron temperature. In order to account for that the heat diffusion equation~\ref{eq:theory_heat_equation_num} was solved numerically using the electron-phonon coupling extracted for large samples (see next section). The resulting curves are shown by the two solid purple lines in Fig.~\ref{fig:supra_hot_electron1}. The two lines originate from the largest and smallest device resistance as this influences the total cooling power through the phonons. The influence for $U\leq$~\SI{1}{\milli\volt} is marginal and cannot account for the observed deviation. In contrast the correction is significant for $U>$~\SI{1}{\milli\volt} for device B and can be as large as \SI{0.8}{\kelvin} for a device resistance of \SI{5.2}{\kilo\ohm} at $U=$~\SI{2}{\milli\volt}. However, the total cooling power through the phonons depends on the device resistance, which is another argument to rule out the phonon cooling as the main origin of the reduced $T_e$ (at small biases) in the first place.

As a last explanation for the reduced $T_e$ at low bias voltages, we propose an increased Lorenz number, which increases the cooling through electron out-diffusion. Even though the Lorenz number $\mathcal{L}_0 = \frac{\pi^2 k_B^2}{3e^2}$ is supposed to be a universal constant, different values have been reported for different materials \cite{1993_Kumar} so far. In order to explain our results, the Lorenz number needs to be increased by \SIrange{24}{48}{\percent}. This is shown by the solid blue lines in Fig.~\ref{fig:supra_hot_electron1}~(b) and (c). For device A we can reproduce the measured electron temperatures by increasing $\mathcal{L}_0$ by \SI{24}{\percent}, whereas for device B and increase by \SI{48}{\percent} is needed.

Previous reports on single layer graphene have also reported an increased Lorenz number between $1.26\mathcal{L}_0$ and $1.34\mathcal{L}_0$ \cite{2013_Fong}. It is theoretically predicted that electron-electron interactions might modify the Lorenz number in graphene \cite{1993_Kumar, 2008_Mueller, 2009_Foster}. It was shown theoretically that in the limit $E_F\ll k_BT$ the system becomes quantum critical and interactions between massless electrons and massless holes increase the Lorenz number \cite{2008_Mueller, 2009_Foster}. However, our samples are clearly not in this regime as $k_BT\ll E_F$ for all temperatures and densities achieved in these experiments. For impurity limited samples, as ours, a modification of the Lorenz number is also expected, but only if screening is weak \cite{2008_Mueller}, which means that the electron-electron interactions are not fully screened.

\subsection{Phonon cooling}
\label{subsec:phonon_cooling}

Sample C has a large area with a length of \SI{100}{\micro\metre}, which promotes phonon cooling over electron out-diffusion. Therefore it is suitable to study the cooling through electron phonon coupling, as the cooling by electron out-diffusion is greatly reduced and a flat temperature profile results, see Fig.~\ref{fig:supra_T_profile}(c).

\begin{figure}[htbp]
	\includegraphics[scale=1]{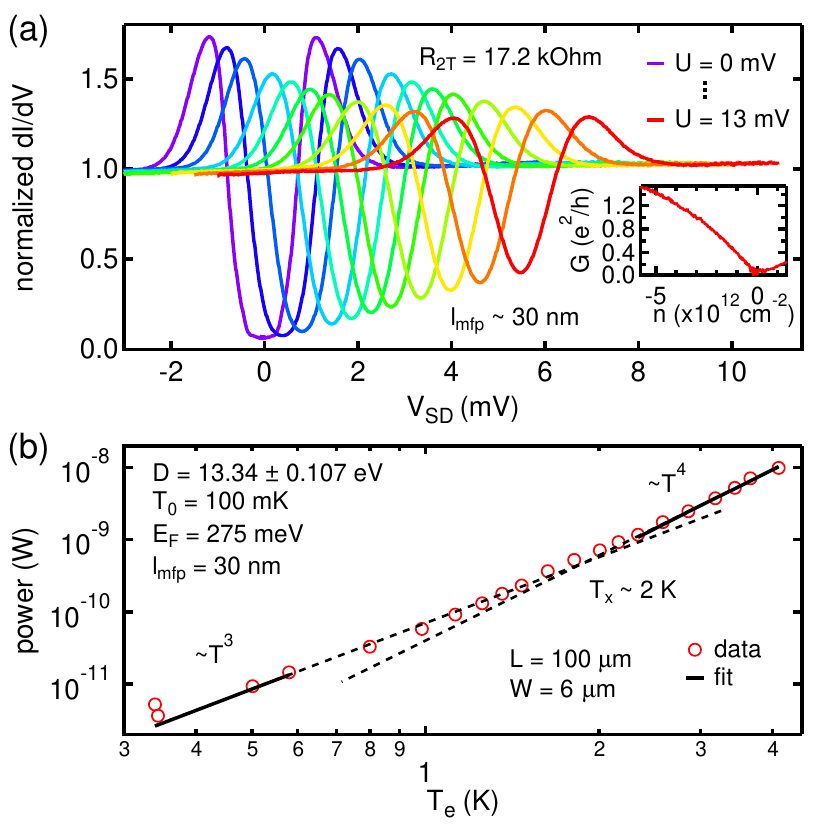}
	\caption{\label{fig:supra_phonon_cooling}\textbf{Device C in the phonon cooled regime:} \textbf{(a)} shows the normalized differential conductance measured through the superconducting electrode to the graphene for different values of heating bias $U$. The inset shows the two-terminal conductance through the graphene from N1 to N2 as a function of density. \textbf{(b)} shows a log-log plot of the calculated Joule heating power versus the extracted electron temperature from fitting a Fermi-Dirac distribution to the numerically extracted energy distribution function. Linear fits with exponent 3 and 4 to the low and high temperature part, respectively, are shown with solid black lines. The dashed lines are guide to the eye showing the transition between the dirty and clean limit around \SI{2}{\kelvin}.}
\end{figure}

The differential conductance traces for device C are shown in Fig.~\ref{fig:supra_phonon_cooling}~(a) for different values of heating bias $U$. All measurements were performed at a high doping of \SI{-5.6e12}{\per\square\centi\metre}. The Joule heating power is shown as a function of the extracted $T_e$ in Fig.~\ref{fig:supra_phonon_cooling}~(b). As seen from Eq.~\ref{eq:theory_heat_equation} the cooling power through the acoustic phonons is described as $P=A\Sigma_{ep}(T_e^\delta - T_0^\delta)$, with the total area A, the electron-phonon coupling $\Sigma_{ep}$, the electron temperature $T_e$ and the phonon temperature $T_0$. In the dirty limit a power law of 3 is expected, whereas it is 4 at higher temperatures in the clean limit.

Since both axes in  Fig.~\ref{fig:supra_phonon_cooling}~(b) are in the log scale, all data points should fall on a line if a single exponent would describe the data over the full temperature range. This is clearly not the case. We also note, that the finite base temperature does not affect this. Whereas the lower temperature points ($T<$\SI{1}{\kelvin}) can be fitted with an exponent $\delta=3$ (dirty limit), the points above \SI{2}{\kelvin} are rather described with an exponent $\delta=4$ (clean limit). This is expected as the dirty limit is more relevant at lower temperatures, which was shown in previous measurements on single layer graphene \cite{2013_Borzenets}.

We have used equation~\ref{eq:theory_heat_phonon_T4} for the high temperature range and and equation~\ref{eq:theory_heat_phonon_T3} for the low temperature range to simultaneously fit the deformation potential $D$. In doing so, we extract $D=$~\SI{13.3}{\eV}. The fits are shown in solid black lines within the fitting range. The extrapolation of the two regimes allows us to extract the cross-over temperature at which the electron-phonon cooling changes from the dirty to the clean limit. This results in a cross-over temperature on the order of \SI{2}{\kelvin}. A crude estimation of the crossover from the dirty to the clean limit by Eq.~\ref{eq:theory_heat_phonon_T_dis} yields a crossover temperature of around \SI{1}{\kelvin} using a mean free path of \SI{\sim 30}{\nano\metre} extracted from the gate dependence of the graphene resistance. This value agrees well with the measurement. 

Similar values for the deformation potential were obtained for a density of \SI{-3.4e12}{\per\square\centi\metre}.

Our extracted value for the deformation potential is within the wide range of literature values that range from \SIrange{2}{70}{\eV} \cite{2012_Fong, 2012_Betz, 2012_Betz_a, 2013_Fong, 2016_McKitterick}. It agrees well with the most reported values around \SI{15}{\eV}, which are in agreement with theoretical predictions ranging from \SIrange{5}{13}{\eV} \cite{2010_Borysenko, 2010_Li, 2012_Kaasbjerg}.

\subsection{Hint of double step function}
\label{subsec:double_step}
\begin{figure}[htbp]
	\includegraphics[scale=1]{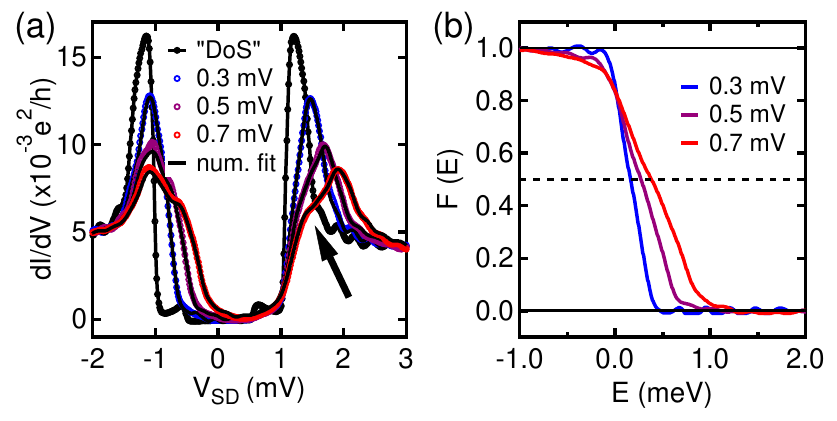}
	\caption{\label{fig:supra_double_Step_hint}\textbf{Hint of a double step function in Device D:} \textbf{(a)} shows the differential conductance measured through the superconducting electrode to the graphene for different values of $U$. Clear shoulders develop for \SI{0.5}{\milli\volt} and \SI{0.7}{\milli\volt}, marked by arrow. The numerically extracted distribution functions are shown in \textbf{(b)}. A hint of a double step with a plateau around 0.5 is visible.}
\end{figure}

The tunnelling differential conductance of Device D is shown in Fig.~\ref{fig:supra_double_Step_hint}~(a). This device developed a shoulder in the conductance peaks at the superconducting gap edges at moderate biasing $U\sim$~\SI{0.5}{\milli\volt}. This shoulder is a first indication of a double step energy distribution function as described in section~\ref{sec:Introduction} for non-interacting quasiparticles. The corresponding numerically extracted distribution functions are shown in Fig.~\ref{fig:supra_double_Step_hint}~(b), and the calculated differential conductance based on these distribution functions reproduce the measured differential conductance very well (thin solid black lines), see Fig.~\ref{fig:supra_double_Step_hint}~(a). Hints of a plateau are visible at 0.5 in the energy distribution functions. However, at larger biases, the energy distribution functions start to smear out due to self heating of the electrons.

There are three limitations present in this data set. First, the differential conductance was only measured within a bias window of \SI{\pm3}{\milli\volt}, that complicates the numerical extraction. Ideally, the differential conductance is measured over a bias range that is much larger than the superconducting gap. Far away from the superconducting gap the differential conductance approaches a constant value that is the normal state conductance. If the differential conductance approaches a constant, it can be numerically extended to any bias value that is optimal for the numerical deconvolution. However, this is not the case here and therefore the deconvolution was performed on a limited bias range. In addition, the measured differential conductance contains some wiggles due to UCF, that were not fully average out (not enough averaging over back gate voltage). The presence of this relatively sharp features that even change with applied bias are a further complication for an accurate extraction of the energy distribution function. A last complication is the additional resonance feature observed within the gap at \SI{\sim 0.5}{\milli\eV}. The exact origin of this is unknown but it might originate from the proximity induced superconducting gap in the Pd layer that was used as a sticking layer. This is a further feature that introduces some complications in the numerical deconvolution and even worse, it might change with bias as well. It was observed that it disappears with increasing temperature and that it is fully absent at \SI{1}{\kelvin} (not shown). Nevertheless, a hint of an additional plateau at 0.5 is observed that is characteristic of non-interacting quasiparticles.

\section{Conclusion}
\label{sec:Conclusion}
In conclusion, superconducting tunnel spectroscopy was successfully used to locally extract the energy distribution function in graphene driven out-of-equilibrium. In the cases where the extracted energy distribution function resembled a simple Fermi-Dirac distribution the local electron temperature was extracted. The dependence of the electron temperature on heating bias or Joule heating power, respectively, revealed a hot electron regime and a phonon cooled regime. The former regime is dominated by electron out-diffusion that is well described by the Wiedemann-Franz law, however with a modified Lorenz number. The increased Lorenz number most probably originates from not fully screened electron-electron interactions. The latter regime is dominated by phonon cooling. In this case, the electron-phonon coupling in the graphene is the bottleneck in cooling hot electrons and we can therefore extract its strength parametrized by the deformation potential $D$.

We have also investigated another sample, sample D, for which we observe signatures of a double-step distribution function originating from non-interacting quasiparticles in the graphene. We believe that a clear double-step distribution function could be observed in samples made from exfoliated graphene encapsulated in exfoliated hBN crystals.

The method presented here can also be used to obtain the density of states if the channel material is kept at equilibrium with a well known distribution function. This has been proven especially powerful for the study of Andreev bound states in carbon nanotubes \cite{2010_Pillet} or graphene \cite{2017_Bretheau}. Therefore, this method could be useful to study band modifications (e.g. graphene minibands or proximity spin-orbit coupling in graphene/TMDC systems) by local measurements of the density of states.



\section*{Acknowledgement}

The authors thank G. F\"ul\"op for helpful discussions. 

This work has received funding from ERC project TopSupra (787414), the European Union Horizon 2020 research and innovation programme under grant agreement No 696656 (Graphene Flagship), the Swiss National Science Foundation, the Swiss Nanoscience Institute, the Swiss NCCR QSIT, Topograph, ISpinText FlagERA networks and from the OTKA FK-123894 grants. P.M. acknowledges support from the Bolyai Fellowship and as a Marie Curie fellow. This research was supported by the National Research, Development and Innovation Fund of Hungary within the Quantum Technology National Excellence Program (Project Nr.  2017-1.2.1-NKP-2017-00001). S.H., Sa.C. and R.W. acknowledge support from the EPSRC (EP/K016636/1, EP/M506485/1).

\section*{Author contribution}
S.Z., P.M. and Se. C. fabricated the devices. The measurements were carried out by S.Z. and P.M. with the help of Se.C. and J.G.. Data was evaluated by S.Z. with help of P.M. and inputs from C.S.. J.G. developed the superconductor deposition. K.T. was growing the CVD graphene used in this study. Sa.C., R.W. and S. H. provided the CVD hBN used for device B and C. The manuscript was written by S.Z. and P.M.. S.Z., P.M., and C.S. designed the experiment. All authors commented on the manuscript.

\newpage
\appendix

\section{Discussion on the sample geometry and finite width of the channel}
\label{sec:app:geometry}
Compared to previous studies with Cu nanowires \cite{1997_Pothier} or CNT \cite{2009_Chen}, we use 2-dimensional graphene as the channel material. We would like to note that the 2D-nature of the device does not influence the predicted temperature profiles derived for a 1D case in the beginning of the manuscript. It is important to note that the sample is translational invariant along the width since the contacts and the channel are assumed to be homogeneous in the direction perpendicular to the transport.

\section{Temperature dependence of the tunneling curves for elevated bath temperatures}
\label{sec:Temp_dependence}

\begin{figure}[htbp]
	\includegraphics[scale=1]{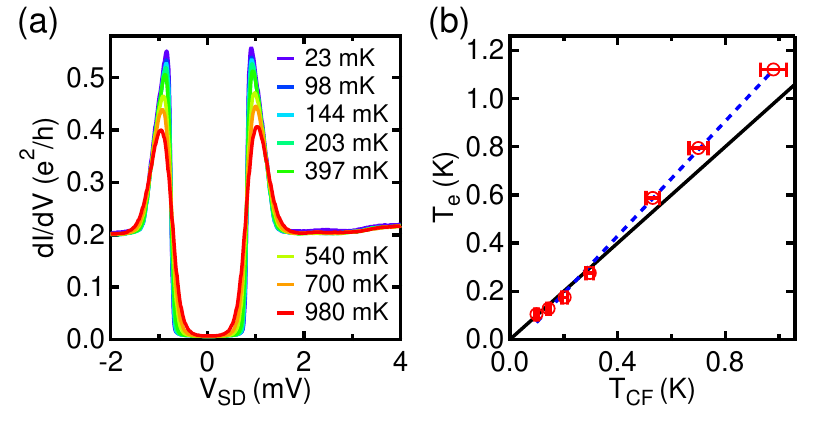}
	\caption{\label{fig:supra_T_dependence}\textbf{Temperature dependence of device A:} \textbf{(a)} shows the differential conductance measured through the superconducting electrode to the graphene for different fridge temperatures with $U =$~\SI{0}{\volt}. \textbf{(b)} The electron temperature $T_e$ extracted from fitting a Fermi-Dirac distribution to the numerically extracted distribution function is shown as a function of the fridge temperature $T_{CF}$. The black line is guide to the eye with slope one, whereas the blue dashed line is a linear guide to the eye through the data points.}
\end{figure}

In order to test the ability of the method presented above to extract the electron temperature, the tunnelling differential conductances was measured at different bath temperatures, which is shown in Fig.~\ref{fig:supra_T_dependence}~(a). An increased bath temperature results in a smearing of the feature in the differential conductance resulting from the superconducting gap. The electron temperature extracted from fitting a Fermi-Dirac distribution to the numerically extracted distribution function is shown in Fig.~\ref{fig:supra_T_dependence}~(b) against the bath temperature measured on the cold finger of the dilution fridge.

A clear linear relation between the bath temperature and the electron temperature is obtained. The electron temperature starts to saturate at low bath temperature and does not decrease further. The negligible change in the tunnel conductance between \SI{23}{\milli\kelvin} and \SI{98}{\milli\kelvin} indicates a lower bound of \SI{\sim 100}{\milli\kelvin} for the electron temperature. It is well known that the electron temperature decouples from the bath temperature if the electrical leads in the fridge are not well thermalized and filtered against high frequency electromagnetic radiation. Even though our set-ups are equipped with RF filters at room temperature and low temperature, a deviation of the electron temperature can still occur.

In addition to the above mentioned deviation at low bath temperature, we face another limitation at low electron temperature. The differential conductance measurement at the base temperature that is used as the DoS of the superconductor contains a finite broadening due to the non-zero electron temperature of this measurement. Therefore, the extracted temperatures close to the base temperature will be underestimated. This effect is negligible at larger temperatures (\SI{\geq 1}{\kelvin}) and will therefore not affect the measurements presented in the main part.

Even though there is a linear relation between $T_e$ and $T_ {CF}$, the electron temperature $T_e$ is always a bit above $T_{CF}$. This could have the following origin. The sample and the thermometer at the cold finger are not exactly at the same position. Furthermore, the thermal coupling of the sample to the cold finger is usually not as good as the one of the thermometer. These two set-up related issues would both lead to $T_e\geq T_{CF}$.

\section{Deconvolution process}
\label{sec:deconvolution}

This section describes the numerical procedure that was used to extract the energy distribution function from the measured differential conductance in more detail. The tunnelling current through a superconductor - insulator - graphene (S/I/Gr) can be written as follows:
\begin{equation}
	\label{eq:app_supra_Tunnel_IV}
	I(V)\propto \int_{-\infty}^{+\infty} \mathrm{d}E n_s(E-eV)n_{gr}(E)\left[f_{gr}(E)-f_s(E-eV)\right].
\end{equation}
The two density of states ($n_s$, $n_{gr}$) and the two energy distribution functions ($f_s$ , $f_{gr}$) determine the current. For small bias values $eV$ on the \si{\milli\eV}-scale, the energy dependence of the graphene density of states can be neglected and assumed to be constant. Therefore, Eq.~\ref{eq:app_supra_Tunnel_IV} can be rewritten
\begin{equation}
	\label{eq:app_supra_Tunnel_IV2}
	\begin{aligned}
	I(V)\propto {}& \int_{-\infty}^{+\infty} \mathrm{d}E n_s(E-eV)\cdot f_{gr}(E)- \\	
	&\int_{-\infty}^{+\infty} \mathrm{d}E n_s(E-eV) \cdot f_s(E-eV),
	\end{aligned}
\end{equation}
where the first integral describes the convolution of the energy distribution function of the graphene $f_{gr}(E)$ with the superconducting density of states $n_s(E-eV)$ and the second integral describes an offset current. The offset current is independent of the bias $V$ and therefore the differential conductance can be written in the following final form:
\begin{equation}
	\label{eq:app_supra_Tunnel_dIdV}
	\frac{\mathrm{d}I(V)}{\mathrm{d}V}\propto \int_{-\infty}^{+\infty} \mathrm{d}E \frac{\mathrm{d}n_s(E-eV)}{\mathrm{d}E}\cdot f_{gr}(E).
\end{equation}
According to Eq.~\ref{eq:app_supra_Tunnel_dIdV}, if $\frac{\mathrm{d}n_s(E-eV)}{\mathrm{d}E}$ is known, then one can use the measured $\frac{\mathrm{d}I(V)}{\mathrm{d}V}$ to extract the energy distribution function in graphene $f_{gr}(E)$ by a deconvolution. There are several ways to perform such a deconvolution: 1) direct deconvolution using built-in algorithms in Matlab, Python etc., 2) Fourier transformation and a simple division, or 3) gradient method, where the distribution function is calculated in many iterations such that the calculated differential conductance fits the measured data. The first and the second method have the drawback of numerical limitations (basically the differential conductance would need to be measured over the whole energy range (-$\infty$ to +$\infty$) with very high accuracy. Since this is not possible, we chose to use the third method: the gradient method, as previously used in similar experiments on copper wires \cite{1997_Pothier}.

The idea behind the gradient method is to start with a reasonable guess of the distribution function and then to calculate the differential conductance based on the guessed distribution function and the known density of states of the superconductor. In our analysis the starting distribution function was a Fermi-Dirac distribution with a guessed electron temperature. The calculated differential conductance is then compared to the measured data and the $\chi^2$ is calculated as defined here:
\begin{equation}
	\label{eq:chi_square}
	\chi^2=\sum_V \left(\left.\frac{\mathrm{d}I}{\mathrm{d}V}\right|^{exp}-\left.\frac{\mathrm{d}I}{\mathrm{d}V}\right|^{calc}\right)^2.
\end{equation}

Now a new distribution function is calculated point by point by adding a small change which is  related to the difference of the guessed and measured differential conductance in the following way:
\begin{equation}
	\label{eq:f-gradient_short}
f_{i+1}(E_k)= f_{i}(E_k)+\lambda\cdot\left.\frac{\mathrm{d}\chi_i^2}{\mathrm{d}f_i(E)}\right|_{E=E_k}.
\end{equation}
Here, $\lambda$ is a small number ($\ll$1) and $\frac{\mathrm{d}\chi_i^2}{\mathrm{d}f_i(E)}$ is the gradient of $\chi_i^2$ with respect to the distribution function evaluated at energy $E_k$ (occupation factor at energy $E_k$) at iteration step $i$. This assures that the distribution function is changed such that the difference between the measured and guessed differential conductance is minimized in the fastest way. Explicitly written, equation \ref{eq:f-gradient_short} reads:
\begin{equation}
	\label{eq:f-gradient_long}
	\begin{aligned}
		 {}& f_{i+1}(E_k)= f_{i}(E_k)+ \\
				& \lambda\sum_V\left[\frac{\mathrm{d}n_s(E-eV)}{\mathrm{d}E}\left(\left.\frac{\mathrm{d}I(V)}{\mathrm{d}V}\right|^{exp}-\left.\frac{\mathrm{d}I(V)}{\mathrm{d}V}\right]^{calc}_i\right)\right]_{E=E_k}.
	\end{aligned}
\end{equation}
The distribution function is updated at every energy $E_k$ with a small change which is a sum over the whole voltage range of the derivative of the density of states multiplied with the deviation of the guessed distribution function from the measured distribution function. In this way, the "non-local" effect of of the convolution in equation \ref{eq:app_supra_Tunnel_dIdV} is reproduced.

%
%

%

\end{document}